# Low emission zones:
# Effects on alternative-fuel vehicle uptake and fleet $CO_2$ emissions

Jens F. Peters [1,*], Mercedes Burguillo[1], José M. Arranz[1]

[1] Universidad de Alcalá, Department of Economics, Alcalá de Henares, Madrid.

* Corresponding author: Jens F. Peters. **Email:** jens.peters@uah.es

**Abstract**

This study analyses the effect of a representative low-emission zone (LEZ) in Madrid on shifting vehicle registrations towards alternative fuel technologies and its effectiveness for reducing vehicle fleet $CO_2$ emissions. Vehicle registration data is combined with real life fuel consumption values on individual vehicle model level, and the impact of the LEZ is then determined via an econometric approach. The increase in alternative fuel vehicles (AFV) registration shares due to the LEZ is found to be significant but fosters rather fossil fuel powered AFV and plug-in hybrid electric vehicles (PHEV) than zero emission vehicles. This is reflected in the average $CO_2$ emissions of newly registered vehicles, which do not decrease significantly. In consequence, while the LEZ is a suitable measure for stimulating the shift towards low emission vehicles, a true zero emission zone would be required for effectively fostering also the decarbonization of the vehicle fleet.

## 1. Introduction

The accelerating climate crisis requires an urgent decarbonization of our economy. For the transport sector, this means, apart from measures for reducing traffic, a quick shift towards alternative fuel vehicles (AFV), powered essentially by renewable electricity (Delbeke, 2016; EEA, 2018; Gnann et al., 2018). In consequence, the uptake of AFV is being fostered by numerous policy measures and incentives like e.g., subsidies, reduced taxes and parking fees or exemption from highway tolls. These (majorly financial) measures were found to have measurable effects on AFV uptake and reducing vehicle fleet emissions, but unlikely to be enough for achieving greenhouse gas (GHG) emission targets in the transport sector (Münzel et al., 2019).

Vehicle purchase decisions are complex and influenced by a multitude of factors. These include technical attributes vehicle like cost, performance, range or fuel savings (Mandys, 2021), the already mentioned policy measures like subsidies, access regulations and charging infrastructure support (Christidis and Focas, 2019a; Homolka et al., 2020; Münzel et al., 2019; Plötz et al., 2017), but also socio-demographic factors like age, education, income or the existence of a second car in the household (Daramy-Williams et al., 2019; Mersky et al., 2016). Apart from these also social influence factors like interpersonal communication, neighbourhood effects and social norms have been found to be important for the decision to purchase an AFV (Jansson et al., 2017; Pettifor et al., 2017). Due to the multitude of factors, it is difficult to clearly determine the effect of an individual parameter on purchase decisions. For instance, while several studies find income to have a significant influence (Singh et al., 2020), others come to the opposite conclusion (Mandys, 2021). Also, the majority of all studies focus on EV uptake, and much less other AFV. A recent review of 239 studies about EV adoption found the highest consensus for the technical variables ‘cost of ownership’ and ‘range’ (rank 2 and 4) and for political measures like development of charging infrastructure (rank 1), purchase incentives / subsidies (rank 3) and government regulations (rank 6) (Kumar and Alok, 2020). Therefore, from a policy viewpoint, a combination of different measures is required for achieving a quick uptake of AFV and the corresponding transition towards low-carbon vehicle fleets (Aklilu, 2020; Bernard and Kichian, 2019; D’Hautfoeille et al., 2016; Plötz et al., 2017). Among these, local conditions (like availability of charging stations, exemption from access restrictions, proximity to bigger cities or household income) seem to have a higher relevance for purchasing decisions than measures on national level like feebate schemes (Christidis and Focas, 2019b; Jiménez et al., 2016; Mersky et al., 2016; Wappelhorst, 2019).

An increasingly popular regional measure for triggering change in vehicle fleet emissions is the introduction of low emission zones (LEZ), restricting access to certain areas of a city according to vehicle emissions (Ezeah et al., 2015; T&E, 2019). While LEZ are usually primarily motivated by concerns regarding air pollution and much less by the need for reducing $CO_2$ emissions (Holman et al., 2015; Morfeld et al., 2011), local access restrictions or LEZ also have a certain effect on renewal of the vehicle fleet, leading to increased replacement of older, non-compliant vehicles by modern low emission vehicles (Ezeah et al., 2015; Quarmby et al., 2019). This, in turn, can be expected to also lower the $CO_2$ emissions of the vehicle fleet. However, studies about the efficacy of LEZ or comparable access restrictions in terms of vehicle purchase decision are rare. For London`s Congestion Charge, the increased replacement rate in London was found to have had a substantial effect on the composition of light duty and passenger vehicle fleet (Ellison et al., 2013) (Morton et al., 2017). Apart from these two studies, to the best of our knowledge no further empirical assessments of the impact of LEZ on AFV uptake have been published so far. Therefore, while a certain influence of such measures on vehicle purchase decisions can be expected (Münzel et al., 2019), the question about their efficacy in terms of reducing fleet $CO_2$ emissions of new vehicles remains widely unanswered. The contribution of the paper is the use of the permanent LEZ of the city of Madrid ('Madrid Central') as an exemplary case study for determining its benefits in terms of fostering AFV uptake and finally decreasing vehicle fleet $CO_2$ emissions, being Madrid Central among the most restrictive LEZ in Europe (Izquierdo et al., 2020; McGrath, 2019; Salas et al., 2021; T&E, 2019). A three-staged approach is applied for this purpose: first (i) determining whether there is evidence that the introduction of the LEZ triggered a significant increase in AFV registrations, then (ii) determining the real life $CO_2$ emissions of all newly registered private passenger vehicles (PPV) and (iii) quantifying the reduction of GHG emissions achieved for newly registered vehicles caused by the introduction of the LEZ. The assessment relies on impact evaluation techniques, estimating an econometric Differences-in-Differences or Double Differences (DID) model using vehicle registration data on individual vehicle level from the Spanish General Traffic Direction and real life fuel consumption values from monitoring portals and surveys. The evaluation methodology is described more in detail in Section 3 of this manuscript, while the underlying data, their origin and handling are explained in Section 4. Previously, Section 2 gives an outline of the policy measures relevant for the assessment, and the outcomes are presented and discussed in Section 5 and 6, respectively. Section 7 provides a summary with policy recommendations.

## 2. Policy framework

In Spain, urban transport policy is mainly driven by concerns regarding air pollution and can be divided into regional and national measures. National measures include vehicle scrapping and feebate schemes aiming at renewing the vehicle park and the introduction of vehicle emission labels. In contrast, regional measures target mainly the access of vehicles to certain areas of the cities depending on the vehicle's emission class. The ones that are most relevant for the local private passenger car transport environment of the two major Spanish cities and that are therefore relevant for the present assessment are depicted in Figure 1 and outlined briefly in the following.

- The **installation of LEZ**, limiting access to the city centre to low emission vehicles.
    - Madrid established a small (~4.7 $km^2$), but very restrictive **permanent LEZ (Madrid Central)** in Dec.2018, limiting access to the LEZ to 'zero' or 'eco' labelled vehicles (details on the vehicle emission labels further below), residents and vehicles accessing one of the municipal parking lots. The idea of establishing a LEZ was first mentioned in the city's air pollution plan (Plan A) end of September 2017, then officially announced in May 2018, and finally established in December 2018. with three month of test phase without sanctioning (Ayuntamiento de Madrid, 2018) (EFE, 2018). As such, it was the first permanent LEZ in Spain and constitutes the reference policy intervention for our assessment.
    - Prior to the permanent LEZ, Madrid established a **non-permanent LEZ** (occasional traffic restrictions on days of high air pollution) in Feb. 2016, following a three-stage protocol (speed limitations, followed by restricted access to the city centre and eventually (stage three) the whole municipal area to vehicles of the highest emission

class (‘A’–labelled vehicles i.e., gasoline registered before the year 2000 and diesel registered before 2006) (Ayuntamiento de Madrid, 2018). Similar to Madrid, in Dec. 2017 Barcelona established access restrictions to older vehicles of the highest emission class (‘A’-label) on days of high air pollution, announcing the measure to become permanent from Jan 2020 on (Ayuntament de Barcelona, 2018).

- Vehicle substitution **subsidy (feebate) schemes**, giving financial help for the purchase of alternative fuel vehicles, but also for the installation of charging infrastructure. These are, except the two subsidy schemes set up by the Region of Madrid in 2019 (Plan MUS), national support schemes and therefore affect all regions in Spain equally. The regional scheme (Plan MUS) only supported vehicle purchases within the province of Madrid and has no equivalent in other regions (CA Madrid, 2019).
- The introduction of national **vehicle emission labels** that classify vehicles into five categories. First for “zero”-labelled vehicles (EV, hydrogen fuelled vehicles and PHEV and with >40km range) in March 2015, and then subsequently for the remaining vehicle classes: “eco” (hybrid electric vehicles and alternative fossil fuel vehicles (AFFV) powered by natural gas (CNG) or liquefied petroleum gas (LPG)) in July 2016 and the three vehicle emission groups A, B, C for conventional cars until 2017. The labels “C”, “B” and “A” classify conventional vehicles according to their emissions, with “C” being Euro 6 compliant, and “A” (most polluting) including all gasoline vehicles registered before 2000 and all diesel before 2006. (DGT, 2016a, 2016b). Although these labels are national and therefore not specific to any city, their introduction (apart from certain awareness-building effect especially in the cities where air pollution is a frequent issue) also brought along several benefits for low polluting vehicles particularly for the two major Spanish cities that regularly suffer from poor air quality (Coves, 2019; Roncero, 2017):
  - Highway toll exemption for EV in Barcelona (BCN)
  - Free parking for all EV i.e., zero -labelled vehicles in restricted parking areas both in Madrid (MAD) and BCN .
  - Reduced vehicle tax (reduction up to 75% for vehicles eco and fully waived for EV) in MAD and BCN.
  - Access to highway express lanes (normally reserved to buses and vehicles with high occupation), both in MAD and BCN for ‘zero’ labelled vehicles.

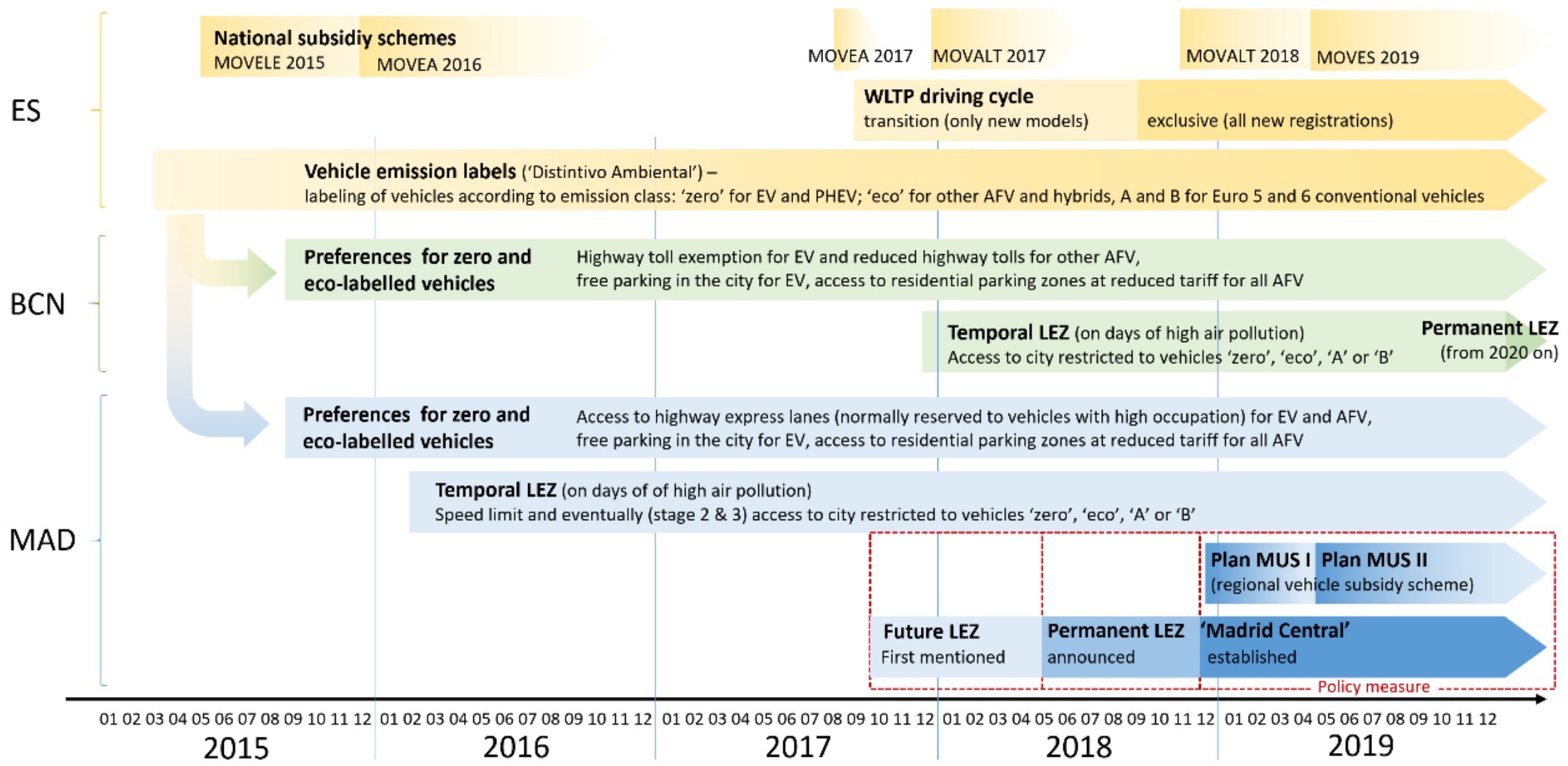

**Figure 1.** Key policy measures introduced in MAD and BCN potentially affecting AFV uptake. The dashed red line marks the policy measures subject to assessment

## 3. Methodology

### 3.1. Estimating the impact of the LEZ on AFV registrations: Econometric model

For assessing the impact of the LEZ in Madrid on AFV registrations, we estimate econometric models to evaluate whether the policy measure (introduction of the LEZ in combination with the regional subsidy scheme) has led to a statistically significant increase in AFV registrations, (Khandker et al., 2009). The principal challenge we are facing is that, apart from the introduction of the LEZ, many factors influence purchase decisions regarding private cars and thus AFV uptake, which are difficult to control for. These are superseded national subsidy programs, but also the simple amount of AFV available on the market, their relative prices compared to equivalent conventional counterparts, available refuelling infrastructure, or specific privileges for AFV like free parking. In addition, also public perception and awareness of air pollution and climate change play relevant roles (Christidis and Focas, 2019a; Mersky et al., 2016; Münzel et al., 2019). While difficult to separate from the effect of the LEZ, these need to be controlled for to avoid biased results. For that purpose, we estimate the potential impact of the introduction of the LEZ in combination with the regional subsidy scheme (as depicted in **Figure 1**) over the share of AFV in total private passenger car registrations within the municipality and province of Madrid using a difference-in-differences (DiD) approach, similar to that applied by Jiménez et al. (Jiménez et al., 2016) for assessing the impact of vehicle subsidize schemes on vehicle prices. A precondition for this is the availability of a counterfactual twin (i.e., a control group not subject to the treatment under assessment) with similar characteristics and parallel trends prior to the treatment. Given that we have a suitable counterfactual control group, the introduction of the LEZ in combination with the regional subsidy scheme can then be considered as a quasi-natural experiment, with the control group allowing to control implicitly for factors that can be assumed to affect both groups equally (like e.g., availability of vehicle models, prices or national incentive schemes) (Albalate, 2008; Khandker et al., 2009). We use Barcelona as control group, being the region most similar to Madrid in terms of socioeconomic parameters and population size. Also, Madrid and Barcelona are the only two Spanish cities where public perception of air quality issues has experienced a significant increase in the recent years and where similar measures for reducing urban air pollution have been established in the years prior to the LEZ (like e.g., parking privileges for AFV, ban of older conventional vehicles from the city centre on days of heavy air pollution; see Figure 1), which likely have a certain influence on consumer choice when buying a new vehicle. The parallel trend of both cities prior to the introduction of the policy measure (Figure 2) is a good indicator for the appropriateness of the control group.

For the assessment we specify three different linear regression models. (Albalate, 2008; Jiménez et al., 2016). The full econometric specification is given by Equation 1, with the dependent variable being the share of AFV (shareAFV, denominating either the total share of all AFV, of AFFV, of PHEV or EV) within total private passenger car registrations between 2015-2019, using monthly aggregated data.

**shareAFV =** **(Equation 1)**

$$\beta0 + \beta1 * \text{treat_MCact} + \beta2 * \text{post_MCact} + \beta3 * \text{DiD_MCact} + \\ + \beta4 * \text{post_MCann} + \beta5 * \text{DiD_MCann} + \beta6 * \text{PlanA} + \beta7 * \text{DiD_PlanA} + \\ + \beta8 * \text{SS_AFFV} + \beta9 * \text{relLPG} + \beta10 * \text{EVchgStat} + \beta11 * \text{TasaAct_} + \varepsilon_t$$

with the following DiD estimators:

- treat_MCact: binary variable that takes the value 1 if the region belongs to the treated group i.e., Madrid *(all Models)*
- post_MCact: binary variable that takes the value 1 for months equal or after Dec 2018, when the LEZ zone (the treatment subject to assessment) was established (all Models)
- DiD_MCact: the difference-in-difference estimator for the introduction of the LEZ,

calculated as the product of treat_MCact and post_MCact *(all Models)*

- post_MCann: binary variable that takes the value 1 for months from May 2018 (when the LEZ zone was officially announced, including concrete conditions and starting date) until Nov 2018 (the month before the actual establishment of the LEZ) *(Model 3)*
- DiD_MCann: the difference-in-difference estimator for the concrete announcement of the LEZ , calculated as the product of treat_MCact and post_MCact *(Model 3)*
- PlanA: binary variable that takes the value 1 for months between oct 2017 and April 2018. This is the period between the first mentioning of a future LEZ in the so-called PlanA of the city of Madrid (yet without specific date) and its official announcement *(Model 3)*
- DiD_PlanA: the difference-in-difference estimator for existence of PlanA, calculated as the product of treat_MCact and post_MCact *(Model 3)*

and the following exogeneous control variables *(Models 2 and 3)*:

- SS_AFFV: CNG and LPG service stations (number of SS that provide LPG or CNG)
- relLPG: relative LPG price (ratio LPG price / price of conventional diesel fuel)
- EVchgStat: Amount of EV charging station available in the province (absolute number*)*
- TasaAct: Activity rate (percentage; value between 0 y 100)

First (**Model 1**), we estimate our model without control variables i.e., $\beta 4 - \beta 11 = 0$

Second (**Model 2**), we add the exogenous control variables $\beta 8 - \beta 11$ to our estimations. A more detailed explanation of these variables and their justification is provided in Section S4 of the SI.

Third (**Model 3**), in order to account for possible announcement effects (purchase decisions might anticipate the actual introduction of the LEZ partially as soon as the LEZ is announced and present in public perception), we incorporate the additional DiD variables $\beta 4$ - $\beta 7$

To further test the econometric model for robustness, we perform an ex-ante ex-post comparison of the summary statistics of the dependent and the descriptive variables, obtaining a small value for the variable-specific DiD indicators. An additional placebo test using the one-year period prior to the LEZ introduction as placebo treatment gives no significant effect for the DiD variable, supporting the appropriateness of the control group. Section S5 of the SI provides further information about the robustness of the econometric models and the suitability of the control group.

### 3.2. Estimating the contribution of the regional subsidy scheme to total AFV registrations

A main difficulty for assessing the effect of the LEZ is its temporal coincidence with a regional subsidy scheme (see Figure 1). The so-called ‘Plan MUS’ assigned a total of 5 Million € to support the purchase of AFV, both personal and light commercial, and the installation of EV charging points in the province of Madrid (CA Madrid, 2019). Being essentially parallel in time, it is impossible to separate the effects of both measures with the available data, which is why we assess them as a package of measures in combination. However, knowing the total amount and the subsidy given for each vehicle type, we can obtain a rough estimation of the effects of this incentive scheme in terms of AFV uptake. Resting these effects from the total AFV shares and running the regressions again with the adjusted values allows to check whether the impact of the policy measure (LEZ) is still significant without the added effect of the subsidy scheme.

The subsidy scheme supported the purchase of AFV with varying amounts depending on the engine type; between 1,500€ for LPG powered vehicles and 5,500€ for EV and PHEV with >70 km electric rage (or EV with range extender)(CA Madrid, 2019). With a total amount of 5 Million € and considering that in 2019 a total 21,000 AFV have been registered within the province, the subsidy scheme can have supported up to between 5 and 15% of the newly registered AFV. However, the scheme also supported light commercial vehicles and installation of EV charging infrastructure. No information is available about the actual distribution of the funds to these different areas, so we assume all the amount to be destined fully to supporting the purchase of private passenger AFV, and that the shares of the different AFV types supported by the plan correspond to the shares of AFV types registered during 2019, the period in which the scheme was active. This approach brings along several uncertainties, since we do not have precise data about the number of vehicles of each AFV type vehicle type and must rely on average values. Also, assuming that all funds are used for exclusively subsidising private passenger vehicles (PPV) will overestimate the effect of the subsidy on AFV uptake, in turn giving a potential underestimation of the isolated effect of the LEZ. In consequence, if we still find significant effects for the policy measure after resting these effects, they will most probably be attributable to the LEZ. The estimated number of vehicles supported under the subsidy scheme is provided in **Table S2** (Section S.3) of the SI.

## 3.3. Quantifying the reduction of fleet $CO_2$ emissions due to increased AFV uptake

### 3.3.1. Real-life $CO_2$ emissions

One of the intended effects of fostering AFV uptake is a reduction of vehicle $CO_2$ emissions. Having determined the effect of the LEZ on vehicle registrations and AFV shares, we now take a closer look at the corresponding impacts on average fleet emissions of newly registered private passenger cars. For this purpose, the real-life fuel consumption values as obtained from surveys and monitoring portals (see section 3.2. for the underlying data and processing approach) are then multiplied with average combustion $CO_2$ emissions of the different motor fuels according to emissions according to EMEP/EEA standards (EEA, 2019). For EV, the direct $CO_2$ emissions associated with the generation of the required electricity are considered using the average $CO_2$ intensity of the Spanish electricity mix for 2019 ( 0.17 $gCO_2/Wh$ (REE, 2020)), but no further upstream emissions from fuel mining and provisioning or transmission losses (since also for combustion vehicles only tailpipe emissions are accounted for, but no emissions from upstream processes like oil drilling, refining, etc.). A summary of the real-life $CO_2$ emission values for each vehicle type is provided in **Table 1**, while a more detailed description of the methodology can be found in Section S.1.2 the SI. Note that these values refer only to newly registered vehicles within the 5-year assessment period (2015-2019), and not the whole vehicle fleet, and that only the direct tailpipe emissions are accounted for, disregarding upstream emissions from fuel extraction and processing. However, the values contain the emissions of over 90% of all newly registered vehicles within the assessment period, giving a comprehensive picture of new vehicle fleet emissions.

### 3.3.2. Reduction of fleet $CO_2$ emissions

For determining the reduction of fleet $CO_2$ emissions caused by the introduction of the LEZ the previously used econometric model is not applicable. We do not find any suitable control group that shows a parallel trend prior to the introduction of the policy measure (and, in consequence, the previously applied DiD approach is not feasible). However, having estimated to potential effect of the introduction of the LEZ on AFV shares, and the $CO_2$ emission values for more than 90% of all vehicle models and the number of AFV registered each month, we can follow a deterministic approach. This allows calculating the reduction of total new PPV fleet $CO_2$ emissions attributable to the share of AFV registered within each month according to Equation2:

$$relCO2Red_{AFV,m} = \left(\frac{CO2_{Rest,m} * sum_{Rest,m}}{sum_{All,m} - sum_{AFV,m}} - CO2_{All,m}\right) * \frac{1}{share_{AFV,m} * 100} \quad (2)$$

*with*

- $relCO2Red_{AFV,m}$: reduction of $CO_2$ emissions [g $CO_2$/km] of total new PPV fleet per percent-point of increase of share of AFV
- $CO2_{Rest,m}$ : average $CO_2$ emissions [g $CO_2$/km] of the remaining newly registered PPV in month m
- $sum_{Rest,m}$: amount of other PPV (excluding the assessed AFV type) registered in month m
- $sum_{All,m}$: total amount of PPV registered in month m
- $sum_{AFV,m}$ : total amount of PPV of the assessed AFV type registered in month m
- $CO2_{All,m}$ : average $CO_2$ emissions [g $CO_2$/km] of all PPV registered in month m
- $share_{AFV,m}$ : assessed AFV type's share of total PPV registrations in month m

In other words, we calculate the average $CO_2$ emissions of the fleet without the assessed AFV type and rest the average emissions of the whole fleet from this (the term in brackets in Equation 2). This provides the reduction in newly registered PPV fleet emissions in the given month due to the assessed AFV type. Dividing the result by the share (percentage) that the assessed AFV type contributed to the total PPV registrations in this month gives an estimation of the reduction per percentage point increase of AFV registration shares (the relation with ß3 from Equation1). The outcome ($relCO2Red_{AFV,m}$) is the average reduction in $CO_2$ emissions of the new PPV fleet achieved by an increase of the share of the given AFV type within all new PPV registrations by one percentage point in month m. Combining this with the effect of the LEZ on increasing the share of each AFV type within total new private passenger vehicles (PPV) registrations as described in **Section 3.1,** the impact of the LEZ in terms of newly registered PPV fleet emissions can be calculated. This requires simply multiplying the estimated coefficient β3 from Equation 1 (the increase in percentage points of the share of the specific AFV due to the introduction of the LEZ) with $relCO2Red_{AFV}$ (the average reduction of $CO_2$ emissions per percent-point of increase of share of the specific AFV).

## 4. Data and descriptive analysis

### 4.1. Vehicle registration data

We rely on public open data sources provided by the city of Madrid (Ayuntamiento de Madrid, 2020), the Spanish General Traffic Direction (DGT) (DGT, 2020) and the Spanish National Statistics Institute (INE) (INE, 2020) for our analysis. DGT provides detailed information about each new vehicle registered every month in Spain, including technical data like weight, engine type and $CO_2$ emissions (according to manufacturer datasheets), plus the exact vehicle type and place of registration from 2015 until 2019. This allows a spatial differentiation of vehicle registrations / the newly registered vehicle fleet on province or municipality level and therefore the evaluation of local impacts. The data is accessible freely from an open data portal but requires substantial post-processing before further use. All data is imported in Stata for statistical analysis. Information about policy measures (e.g., subsidies of traffic restrictions within the LEZ or dates and duration of vehicle subsidy programs) are collected from various sources, including newspaper articles and official websites and information portals. Fuel prices and macroeconomic data (employment, income) is taken from INE (INE, 2020), the amount of EV charging points available in each region is provided as a courtesy by Electromaps (Electromaps, 2020). More details about data sources and post-processing are provided in **Section S1**. of the Online Supplementary Information (SI), while **Table 1** provides the summary statistics for the relevant variables. We limit our analysis to private vehicles, since these make up the major share of new passenger car registrations (82.5%, followed by 16.5% rental cars (including sharing), while all others, including Taxis and comparable chauffeur vehicles (e.g., Uber, Cabify etc.), contribute only about 1%) (DGT, 2020). We explicitly exclude all sharing and renting vehicles, being the parameter of interest the impact of the LEZ on individual vehicle purchase decisions, which follow other criteria than those of rental or sharing car companies. **Figure 2** shows the share of AFV in total new private passenger vehicles (PPV) registered per month in Madrid (MAD) and Barcelona (BCN), both on municipality as on province level. The trends are strikingly similar for Madrid and Barcelona prior to the first announcement of the LEZ, supporting the use of the latter as control group. Fluctuations in AFV registration shares can be attributable to some extent to the (non-) availability of national vehicle subsidy schemes (compare Figure 1), temporarily boosting AFV sales in both Madrid and the control region. Graphs for the individual AFV types (PHEV, AFFV and EV) and more a more detailed discussion of these effects are provided in **Section S2** of the SI.

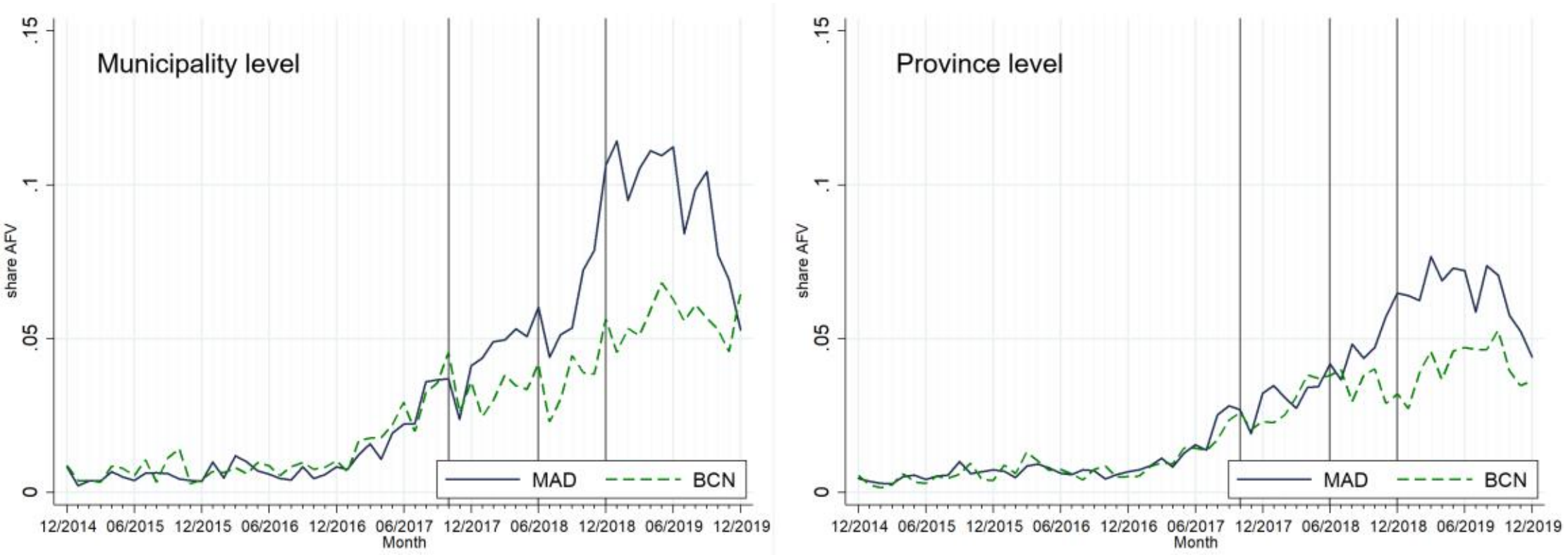

**Figure 2.** AFV registration shares in the provinces (left) and municipalities (right) of Madrid and Barcelona (control group). MAD = Madrid i.e., the city where the LEZ was introduced, BCN = Barcelona, the control city. The vertical bars mark the three key dates of the LEZ introduction: its first mentioning in the public debate, its official announcement, and its effective beginning

### 4.2. Real life $CO_2$ emission data

Although $CO_2$ emissions are provided for each vehicle type in the vehicle registration database, these are manufacturer datasheet values that do not reflect actual $CO_2$ emissions (Helmers et al., 2019; Tietge et al., 2019). For determining $CO_2$ emissions, we therefore rely on data for real fuel consumption obtained manually from various fuel monitoring portals and surveys (DLR, 2005; Fisch und Fischl GmbH, 2017; Travelcard, 2020). Here, vehicle owners introduce the amounts of fuel filled at each refuelling event, together with the km driven on a voluntary basis, allowing for obtaining actual consumption values instead of theoretical driving cycle values. These certainly depend on the driving behaviour of each individual vehicle owner, but due to the high amount of entries in the databases the average values can be expected to fairly reflect the actual reality. Also, real life fuel consumption values determined by different monitoring systems following different approaches (logbook, voluntary database, fuel cards) and for different countries (Germany, France, Spain, Netherlands) were found to be very similar (Tietge et al., 2019; Travelcard, 2020) and do not seem to vary fundamentally between EU countries (Tietge et al., 2019). They can therefore be considered a valid and realistic approximation of real fuel consumption and, in consequence, $CO_2$ emissions for our assessment. For all vehicles with combustion engines, only the direct tailpipe emissions are accounted for, disregarding upstream emissions from fuel extraction and processing. Similarly, for EV the direct $CO_2$ emissions associated with the generation of the required electricity is considered using the average $CO_2$ intensity of the Spanish electricity mix for 2019, but no further upstream emissions from fuel extraction and provisioning or transmission losses. For PHEV, however, only the direct emissions from fuel combustion are included, and not the emissions from electricity. This is because for PHEV no reliable data about electricity consumption can be found (vehicle owners seem to either protocol their fuel or their electricity consumption, but rarely both). A recent study found the real-life utility factor (i.e., the share of electric driving) to be roughly half of that assumed in the driving cycle (Plötz et al., 2020). Including this would increase the actual $CO_2$ emissions by around 10% (see Section S.1.2 of the SI for more details). However, we prefer not to mix methodologies (reported actual fuel consumption versus electricity demand estimated from driving cycle values) and to omit the share of emissions associated with electricity generation. In consequence, it has to be considered that the actual emissions of PHEV tend to be underestimated in the following assessment.

In order to have reasonable amounts of data per vehicle type we aggregate several model years and engine power classes, obtaining six engine types (Gasoline, Diesel, CNG, LPG, PHEV, EV) and five engine power classes (0-74kW, 75-89kW, 90-111kW, 112-148kW, 149-222kW, > 223kW) per vehicle model. In summary, we obtain a fuel consumption database with entries for 2,697 car models, allowing us to attribute real fuel consumption values to 5,066,267 of 5,606,939 entries in our vehicle registration database, thus covering 90.4% of all private passenger car registrations. The corresponding real-life emissions determined for all individual vehicle models are resumed in Table 1 as summary statistic and provided as a data file in the supplementary information. Note that the values in Table 1 are averages over the whole assessment period, while actual fleet emission values change (increase) slightly over time. The corresponding time series are provided in Figures S3-S4 (driving cycle emissions) and Figures S5-S6 (real driving emissions) of the SI.

**Table 1.** Summary statistics for the different vehicle types for the two provinces under assessment. *: EV emissions are calculated based on the electricity demand and average $CO_2$ intensity of the electricity mix

| | | Madrid province | | | | | Barcelona province | | | | | |
|---|---|---|---|---|---|---|---|---|---|---|---|---|
| **EngType** | **Variable** | **Obs** | **Mean** | **Std.Dev** | **Min** | **Max** | **Obs** | **Mean** | **Std.Dev** | **Min** | **Max** | **Unit** |
| **Gasoline** | **Weight** | 541,114 | 1,319.0 | 223.42 | 14.0 | 3,750.0 | 383,180 | 1,286.3 | 205.65 | 14.0 | 3,750.0 | **kg** |
| | **EngPower** | 540,411 | 91.3 | 40.41 | 1.0 | 998.0 | 383,124 | 89.5 | 38.08 | 1.1 | 998.0 | **kW** |
| | **CO2_real** | 522,006 | 168.1 | 28.00 | 13.1 | 533.4 | 371,807 | 169.0 | 25.90 | 108.0 | 533.4 | **$gCO_2$/km** |
| | **CO2** | 539,121 | 120.4 | 25.45 | 27.0 | 545.0 | 381,899 | 122.3 | 23.80 | 27.0 | 591.0 | **$gCO_2$/km** |
| **Diesel** | **Weight** | 836,655 | 1,505.5 | 240.77 | 2.0 | 3,500.0 | 323,235 | 1,511.0 | 247.02 | 4.0 | 3,240.0 | **kg** |
| | **EngPower** | 836,115 | 97.8 | 30.26 | 1.0 | 995.0 | 323,132 | 97.8 | 28.92 | 1.0 | 990.0 | **kW** |
| | **CO2_real** | 793,774 | 161.6 | 27.08 | 89.5 | 391.7 | 302,879 | 163.4 | 27.19 | 89.5 | 442.4 | **$gCO_2$/km** |
| | **CO2** | 835,768 | 116.1 | 22.27 | 44.0 | 365.0 | 322,187 | 119.0 | 23.87 | 40.0 | 389.0 | **$gCO_2$/km** |
| **LPG** | **Weight** | 16,412 | 1,255.8 | 129.70 | 444.0 | 2,546.0 | 5,136 | 1,281.0 | 123.04 | 137.0 | 2,180.0 | **kg** |
| | **EngPower** | 16,405 | 72.8 | 14.67 | 10.7 | 269.0 | 5,136 | 75.2 | 13.71 | 35.8 | 177.0 | **kW** |
| | **CO2_real** | 16,086 | 135.6 | 12.81 | 112.1 | 279.3 | 4,980 | 138.4 | 13.65 | 112.1 | 189.4 | **$gCO_2$/km** |
| | **CO2** | 16,405 | 122.7 | 15.21 | 89.0 | 398.0 | 5,134 | 126.4 | 15.18 | 86.0 | 312.0 | **$gCO_2$/km** |
| **CNG** | **Weight** | 5,258 | 1,329.6 | 101.54 | 1,031.0 | 1,950.0 | 2,311 | 1,324.7 | 97.63 | 1,031.0 | 1,743.0 | **kg** |
| | **EngPower** | 5,257 | 79.8 | 13.73 | 44.0 | 125.0 | 2,311 | 77.4 | 12.73 | 50.0 | 125.0 | **kW** |
| | **CO2_real** | 5,123 | 110.3 | 8.30 | 3.0 | 152.6 | 2,288 | 109.4 | 7.19 | 94.2 | 152.6 | **$gCO_2$/km** |
| | **CO2** | 5,251 | 96.1 | 7.56 | 79.0 | 194.0 | 2,307 | 95.5 | 6.82 | 79.0 | 167.0 | **$gCO_2$/km** |
| **PHEV** | **Weight** | 11,400 | 1,917.3 | 294.88 | 17.0 | 2,751.0 | 2,302 | 1,883.3 | 287.72 | 870.0 | 2,751.0 | **kg** |
| | **EngPower** | 11,397 | 139.6 | 59.53 | 11.0 | 447.0 | 2,302 | 129.1 | 57.25 | 25.0 | 404.0 | **kW** |
| | **CO2_real** | 9,857 | 119.6 | 45.10 | 11.9 | 253.9 | 2,013 | 111.3 | 46.72 | 11.9 | 222.3 | **$gCO_2$/km** |
| | **CO2** | 11,199 | 47.1 | 14.47 | 10.0 | 193.0 | 2,284 | 45.9 | 16.26 | 12.0 | 195.0 | **$gCO_2$/km** |
| | **Electr.Cons** | 11,203 | 144.42 | 33.6 | 13 | 650 | 2,229 | 138.9 | 32.455 | 18 | 240 | **Wh/km** |
| **Electric** | **Weight** | 9,211 | 1,535.2 | 305.82 | 1.0 | 2,960.0 | 4,644 | 1,664.7 | 359.56 | 787.0 | 2,960.0 | **kg** |
| | **EngPower** | 8,198 | 100.6 | 98.60 | 1.0 | 599.0 | 4,389 | 136.4 | 130.61 | 15.0 | 599.0 | **kW** |
| | **CO2_real*** | 9,146 | 27.3 | 4.60 | 16.6 | 92.1 | 4,607 | 28.8 | 5.94 | 16.6 | 72.4 | **$gCO_2$/km** |
| | **CO2*** | 9,074 | 26.5 | 5.25 | 17.0 | 92.1 | 4,550 | 28.8 | 6.44 | 17.0 | 83.5 | **$gCO_2$/km** |
| | **Electr.Cons** | 9,074 | 156.06 | 30.86 | 100 | 542 | 4,550 | 169.2 | 37.86 | 100 | 491 | **Wh/km** |

## 5. Results

### 5.1. Impact of the introduction of the LEZ in combination with subsidy scheme on AFV uptake

#### Province level

The econometric regressions (**Table A3**; results tables are provided in the Appendix due to their size) yield for all three models that the increase of AFV registrations is significant on a 1% confidence level. The estimated effect of the introduction of the LEZ in combination with the regional subsidy scheme (the policy measure 'package') is an increase in the share of AFV within total private passenger vehicle registrations by between 2.2 and 2.3 percentage points (variable 'DiD_MCact' in **Table A3**). However, when breaking this down to the different AFV types i.e., testing for the individual AFV shares using the same equation, the increase is significant only for PHEV (increase by 1.0 - 1.2 percentage points) and AFFV (1.1-1.3 points), but not for EV. A significant influence factor is the activity rate, with a negative contribution except for AFFV, and the amount of EV charging stations available. An announcement effect can also be stated for AFV in general (0.9 points) and PHEV (0.5 and 0.35 points; Model 3, variable 'DiD_MCann' and 'DiD_PlanA' in **Table A3**). The estimated size of the effect does not vary strongly between the different regressions, indicating a robust regression model (more tests in this regard are provided in Section S5 of the SI). Therefore, we can state that on province level the introduction of the LEZ in combination with the regional subsidy scheme triggered a significant increase of AFV sales. This affected majorly PHEV and AFFV while for EV the differences to the control group are statistically not significant.

#### Municipality level

As expected, on municipality level the estimated effect of the introduction of the LEZ in combination with the regional subsidy scheme (the policy measure 'package') is stronger than on province level. The increase in the share of AFV within total private passenger vehicle registrations is between 3.5 and 3.9 percentage points (at 1% significance level; variable 'DiD_MCact' in **Table A4** of the Appendix). Announcement effects are found for AFV (1.9 points), AFFV (1.5 points) and PHEV (0.6 points), but not EV. In addition, also the first mention of the intention to establish a LEZ (variable 'DiDPlanA') caused a increase in AFV and AFFV registrations, though at lower significance level (10%). Unlike on province level, on municipality level the highest effect is found for AFFV (between 3.6 and 3.9 percentage points) and less for PHEV (0.9 - 1.0 points), all at 1% significance level. Interestingly, for EV no increase but a significant reduction of EV shares is found, between -0.9 and 1.0 points. This indicates that the increase in AFFV and PHEV registrations was partially on expense of EV registration. Again, activity rate and charging station availability are significant parameter for all AFV types except for EV, while the amount of AFV service stations shows a significant effect only for AFFV and AFV in general.

### 5.2. Isolated impact of the introduction of the LEZ on AFV uptake

For being able to assess the isolated effect of the LEZ, the number of vehicles supported under the regional subsidy scheme is estimated and rested these from the total number of AFV registered every month (see Section 3.2 for details). The hypothesis is then tested again with the corrected AFV shares, using the same econometric model as described previously. The results are provided in **Tables A5 and A6** of the Appendix.

The effect of the LEZ is still significant with all three regression models, but with slightly lower coefficients. On **province level**, the estimated effect if the LEZ is an increase of AFV shares of between 1.7 and 1.8 percentage points, of AFFV share between 0.9 and 1.0 points, and PHEV share of 0.9-1.1 points, and no significant effects for EV. On **municipality level**, the corresponding values are an increase in AFV registrations due to the LEZ by between 2.4 and 3.1 percentage points, for AFFV between 2.9 and 3.4 points, and for PHEV by 0.7-0.9 points, again, all on 1% significance level. The reduction of EV shares is estimated to be between 1.0 and 1.1 points. **Figure 3** summarizes the obtained coefficients graphically, with the bars indicating the (lower end) estimations from Model 3, and the error whiskers the range of coefficient values obtained by all three models. As mentioned, the present approach tends to overestimate the effect of the local subsidy scheme rather than underestimating it. In consequence, when the impact of the LEZ is still found to be significant after resting the estimated effect of the local subsidy scheme, this can be considered a robust estimation, being the true isolated effect potentially even higher.

In consequence, we can state that the introduction of the LEZ on its own triggered a significant increase in AFV registrations in the province and municipality of Madrid, also when resting the potential effects of the parallel subsidy scheme. The LEZ therefore not only has effects on vehicle purchase decisions in its immediate radius (municipality), but also in the wider surroundings (province). However, it did not have any positive effect on EV uptake, which, on municipality level, was significantly lower than that of the control group, leading to a significant negative coefficient.

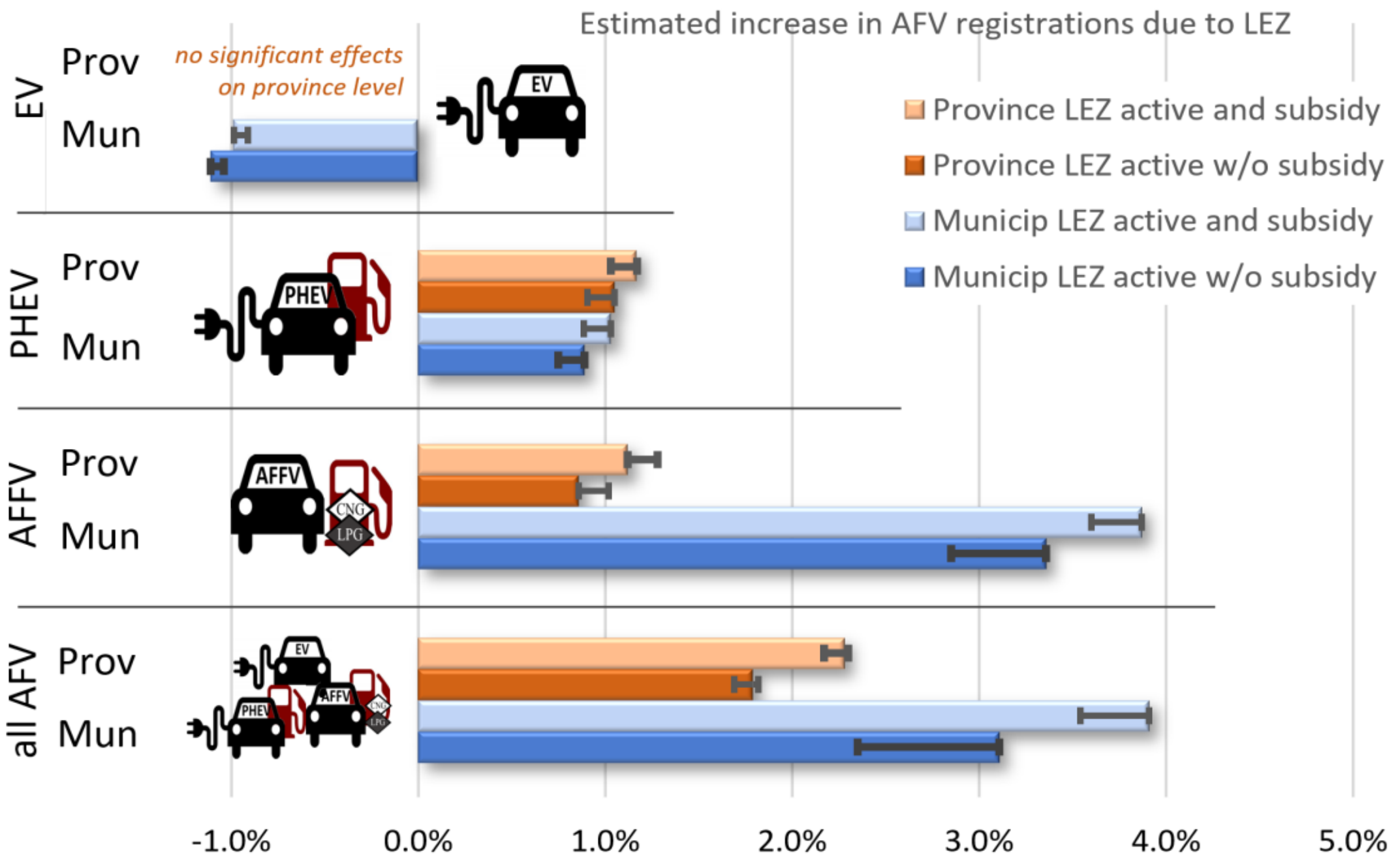

**Figure 3**. Effects of the LEZ in terms of increasing shares of the different types of AFV within total vehicle registrations. Bars indicate results obtained with regression Model 3, and the whiskers display the range between the minimum and the maximum coefficient value obtained from the three regression models. The underlying data is provided in Tables A3 – A6 of the Appendix.

### 5.3. Impact of increasing AFV shares on $CO_2$ emissions

Having determined the real $CO_2$ emissions of all relevant vehicle models registered within the assessment period, we are now interested in knowing whether the reduction of $CO_2$ emissions is significant, what is its magnitude, and to which type of AFV it can be attributed majorly. Unfortunately, the previous DiD approach is not applicable, showing none of the possible control groups a similar behaviour to Madrid in terms of fleet $CO_2$ emissions i.e., a parallel trend prior to the policy measure (see **Section S6.2** of the SI for more details). We therefore use the deterministic approach described in **Section 3.3.2** for calculating the reduction of total fleet $CO_2$ emissions attributable to increasing share of AFV registrations within each month. The five-year average is a reduction in $CO_2$ emissions per percentage point of increase in share of the corresponding AFV type of 0.33-0.39 g $CO_2$/km for AFFV, 0.44-0-54 g $CO_2$/km for PHEV and 1.36-1.37 g $CO_2$/km for EV. Note that for PHEV the emissions from electricity are disregarded (unlike EV, whose benefit would increase to 1.65 g $CO_2$/km if the $CO_2$ emissions of electricity generation were neglected) and would add an estimated 10% to the total $CO_2$ emissions (see Section 3.2.). Actual benefits of PHEV are therefore overestimated and will actually be lower.

### 5.4. Effect of LEZ on reducing vehicle fleet $CO_2$ emissions

With the estimated effect of the LEZ on each AFV type as determined previously (coefficient ß3 in Equation 1) and their corresponding emissions reduction potential, the total reduction in $CO_2$ emissions for the newly registered PPV fleet achieved by the introduction of the LEZ can then be determined. Figure 4 provides the results both with and without the effect of the local subsidy scheme on province and municipality level.,

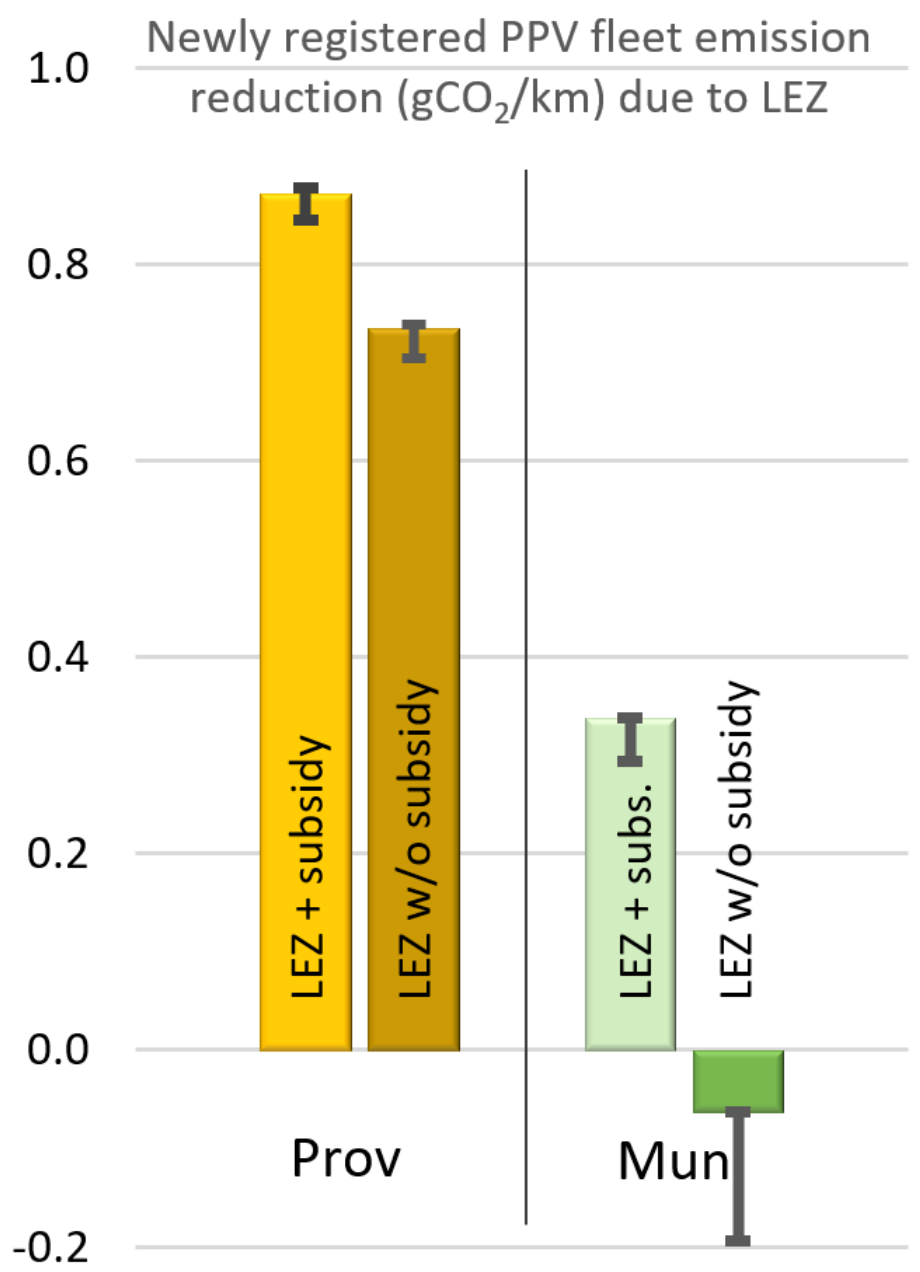

**Figure 4.** Total reduction in $CO_2$ emissions for the newly registered private passenger vehicle (PPV) fleet achieved estimated for the introduction of the LEZ in combination with the subsidy scheme and without on province and municipality level

On **province level**, a rather marginal reduction of between 0.86-0.88 g $CO_2$/km for the fleet of newly registered PPV is obtained for the combination of LEZ and subsidy scheme, and of 0.70-0.74 for the LEZ alone. Of those, 0.37-0.42 g $CO_2$/km can be attributed to the increasing share of AFFV, and 0.44-0.50 g $CO_2$/km due to PHEV (0.28-0.34 and 0.39-0.45 g $CO_2$/km for the LEZ without subsidy, respectively). Since for EV no significant effect is estimated for the LEZ (Figure 3), they do not contribute to the emission reduction effects.

On **municipality level**, the estimated effect shows a higher variation. Depending on the applied regression model, the introduction of the LEZ caused a reduction of only between 0.29 and 0.34 $gCO_2$/km for the combination of both measures and an increase of 0.06-0.19 $gCO_2$/km for the LEZ alone. This is a direct effect of the estimated decrease of EV shares relative to the control group, found to be significant with all three regression models. On municipal level, the increase in AFFV shares causes a reduction of new PPV fleet $CO_2$ emissions by between 1.19 and 1.28 $CO_2$/km, PHEV between 0.38 and 0.44 $CO_2$/km, while for EV, their dropping share within total PPV registrations in combination with their high $CO_2$ emissions reduction potential leads to an increase of emissions between 1.27 and 1.38 g $CO_2$/km (reductions of 0.94-1.11, 0.32-0-38 and increase of 1.47-1.55 g $CO_2$/km for the LEZ alone, respectively). More details about the CO2 reduction potential of the different AFV types can be found in Section S7 of the SI.

## 6. Discussion

### 6.1. Target achievement

The national decarbonization plan sets a target of zero $CO_2$ emissions for newly registered vehicles in 2040 (PNIEC, 2020). However, the current trends clearly point into the opposite direction and are far from reaching this target. Similarly, the also ambitious target of 5Mio EV in 2030 is out of reach under current support policy, requiring a rapid and decided change of this trend. Previous studies found subsidy schemes alone to be insufficient for fostering the required change in vehicle fleet composition, requiring a combination of different measures, including local access restrictions for achieving the transition towards zero emission mobility (Münzel et al., 2019; Wappelhorst, 2019). We find the combination of subsidy schemes with access restrictions as applied in the city of Madrid to effectively contribute to accelerating AFV uptake, but not so for reducing the $CO_2$ emissions of the vehicle fleet. This is because the major effects in terms of increased registrations are found for AFFV and PHEV, which show very limited $CO_2$ reduction potential. In fact, their real driving $CO_2$ emissions are only modestly below those of conventional vehicles (Table 1). EV on the other hand, which show by far the highest potential for reducing $CO_2$ emissions, perceive no significant increase or even decrease in comparison with the control group.

For PHEV in particular, the actual $CO_2$ emission reduction potential is substantially lower than it would be expected from the datasheet values. This can be attributed to two effects: First, PHEV show a particularly large gap between real life $CO_2$ emissions and the driving-cycle values, with an average gap of 250% compared to 140% for conventional gasoline and diesel vehicles and 110-115% for AFFV (see **Table 1** and **Figures S7** / S8 of the SI). Second, PHEV are substantially bigger and equipped with larger engines than other vehicle types, weighing more than 300kg more than the average diesel and 600kg more than the average gasoline car and showing between 43-53% more engine power (average for the province of Madrid, see **Table 1**). In this sense, the unconditional support of PHEV seems to jeopardise the fulfilment of $CO_2$ reduction targets, beneficiating above all heavy upper segment vehicles with misleading emission values. A more differentiated support limited to small PHEV with capped weight and engine power could help to stop this trend and to convert PHEV into an attractive low carbon solution especially in urban environments. Also, a better and broader monitoring of real driving fuel and electricity consumption would be helpful for obtaining more realistic actual values in this regard and would allow to include also the additional electricity demand and corresponding upstream emissions.

## 7. Conclusions and policy implications

Local policy measures like low or zero emission zones (LEZ or ZEZ) are motivated mainly by concerns about air pollution and public health due to contamination. However, since they target traffic and vehicle fleet composition, they are also relevant for decarbonizing the transport sector, and might contribute to also tackle climate change. An assessment of the efficacy of these measures in terms of greenhouse gas (GHG) reduction is therefore highly relevant. We find the LEZ in Madrid to be an effective measure for triggering change, having had a significant impact on alternative fuel vehicle (AFV) registrations in the city and its surroundings. However, it needs to be designed properly. While the admission of alternative fossil fuel cars seems to be reasonable under air pollution aspects, it gives the wrong incentives in terms of decarbonization of the vehicle fleet, shifting purchase decisions towards alternative fossil fuel powered vehicles (AFFV) and plug-in hybrid electric vehicles (PHEV), both majorly fossil fuel powered and therefore with only limited $CO_2$ reduction potential. In fact, the noteworthy impact on increasing AFV shares does not lead to significantly lower $CO_2$ emissions of the newly registered vehicles, making the LEZ ineffective in terms of decreasing the vehicle fleet's GHG emissions. EV (which are the single vehicle type with substantial potential for reducing $CO_2$ fleet emissions) are the only AFV that do not receive a significant push by the introduced measure, and even decrease relative to the control group. Thus, while the LEZ does have potential for shifting consumer choice towards alternative vehicles, it needs to be set up forward-looking, targeting the 2040 objective of 100% zero-emission vehicle registrations: Strict zero emission zones (ZEZ) might be more efficient for achieving the required quick transition in this sense. Apart from that, it must be considered that the present assessment focuses on fleet emissions, and disregards any impacts in traffic volume (e.g., due to modal shift or avoided trips), which might be at least as relevant.

**Glossary**

| | |
|---|---|
| PPV | Private passenger vehicle |
| AFV | Alternative fuel vehicle |
| AFFV | Alternative fossil fuel vehicle (CNG or LPG powered) |
| CNG | Compressed natural gas |
| DiD | Difference in differences (econometric approach) |
| EV | Electric vehicle |
| GHG | Greenhouse gas |
| LPG | Liquefied petroleum gas |
| PHEV | Plug-In hybrid electric vehicle |
| LEZ | Low emission zone |
| ZEZ | Zero emission zone |

## Appendix

**Table A1**. Summary statistics for the dependent variables (monthly number of vehicle registrations of each vehicle type). The shares of each vehicle type is calculated from the total monthly vehicle registrations

| Variable | Obs | Mean | Std.Dev | Min | Max |
|---|---|---|---|---|---|
| *MAD Municipality* | | | | | |
| sumGasl | 61 | 2739.7 | 635.0 | 1371 | 4419 |
| sumDies | 61 | 1908.6 | 705.8 | 721 | 3672 |
| sumLPG | 60 | 79.8 | 97.0 | 1 | 316 |
| sumCNG | 51 | 26.9 | 29.2 | 1 | 102 |
| sumPHEV | 61 | 38.0 | 29.7 | 3 | 144 |
| sumEV | 61 | 35.8 | 33.1 | 1 | 179 |
| *BCN Municipality* | | | | | |
| sumGasl | 61 | 1260.2 | 335.6 | 611 | 1976 |
| sumDies | 61 | 723.4 | 284.3 | 293 | 1502 |
| sumLPG | 57 | 13.5 | 13.1 | 1 | 49 |
| sumCNG | 39 | 7.6 | 6.5 | 1 | 20 |
| sumPHEV | 57 | 10.5 | 5.5 | 1 | 23 |
| sumEV | 61 | 26.3 | 22.0 | 1 | 105 |
| *MAD Province* | | | | | |
| sumGasl | 61 | 8888.3 | 3731.8 | 3333 | 16802 |
| sumDies | 61 | 13730.6 | 2157.8 | 8601 | 17660 |
| sumLPG | 61 | 269.1 | 320.9 | 8 | 1043 |
| sumCNG | 61 | 86.2 | 106.4 | 1 | 375 |
| sumPHEV | 61 | 186.9 | 173.6 | 6 | 684 |
| sumEV | 61 | 151.0 | 143.2 | 5 | 594 |
| *BCN Province* | | | | | |
| sumGasl | 61 | 6287.8 | 1875.9 | 2595 | 10408 |
| sumDies | 61 | 5303.8 | 2226.7 | 1895 | 13004 |
| sumLPG | 61 | 84.2 | 97.6 | 2 | 395 |
| sumCNG | 58 | 39.8 | 39.9 | 1 | 126 |
| sumPHEV | 61 | 37.7 | 19.9 | 5 | 92 |
| sumEV | 61 | 76.1 | 61.1 | 8 | 300 |

**Table A2**. Summary statistics for the independent (control) variables. SS_AFFV: Alternative fuel service stations (CNG and LPG); relLPG: price of LPG fuel relative to conventional diesel fuel; relCBG: price of CNG fuel relative to conventional diesel; EVchgStat: Number of available EV charging stations; PIBcap: per-capita GDP; TasaAct: Activity rate (employment)

| Variable | Obs | Mean | Std.Dev | Min | Max |
|---|---|---|---|---|---|
| *MAD Municipality* | | | | | |
| SS_AFFV | 61 | 71.1 | 9.8 | 56 | 93 |
| relLPG | 61 | 0.6 | 0.0 | 0.537 | 0.618 |
| EVchgStat | 61 | 221.0 | 119.7 | 94 | 503 |
| PIBcap_ | 61 | 37779.6 | 1564.2 | 34837 | 40470 |
| TasaAct_ | 61 | 63.4 | 0.8 | 62.37 | 65.06 |
| *BCN Municipality* | | | | | |
| SS_AFFV | 61 | 107.7 | 11.2 | 92 | 124 |
| relLPG | 61 | 0.6 | 0.0 | 0.52 | 0.605 |
| EVchgStat | 61 | 378.9 | 151.8 | 184 | 683 |
| PIBcap_ | 61 | 36416.5 | 1655.9 | 33430 | 38980 |
| TasaAct_ | 61 | 61.9 | 0.4 | 61.02 | 62.57 |
| *MAD Province* | | | | | |
| SS_AFFV | 61 | 71.1 | 9.8 | 56 | 93 |
| relLPG | 61 | 0.6 | 0.0 | 0.537 | 0.618 |
| EVchgStat | 61 | 221.0 | 119.7 | 94 | 503 |
| PIBcap_ | 61 | 33425.3 | 1596.2 | 30574 | 36041 |
| TasaAct_ | 61 | 63.4 | 0.8 | 62.37 | 65.06 |
| *BCN Province* | | | | | |
| SS_AFFV | 61 | 107.7 | 11.2 | 92 | 124 |
| relLPG | 61 | 0.6 | 0.0 | 0.52 | 0.605 |
| EVchgStat | 61 | 378.9 | 151.8 | 184 | 683 |
| PIBcap_ | 61 | 29323.7 | 1636.7 | 26491 | 32067 |
| TasaAct_ | 61 | 61.9 | 0.4 | 61.02 | 62.57 |

**Table A3.** Regression results on province level, LEZ in combination with local subsidy scheme. Standard errors in parentheses. $^{*}$ $p < 0.1$, $^{**}$ $p < 0.05$, $^{***}$ $p < 0.01$

| **Province** | **(Model 1)** | **(Model 2) shareAFV** | **(Model 3)** | **(Model 1)** | **(Model 2) shareAFFV** | **(Model 3)** | **(Model 1)** | **(Model 2) sharePHEV** | **(Model 3)** | **(Model 1)** | **(Model 2) shareEV** | **(Model 3)** |
|---|---|---|---|---|---|---|---|---|---|---|---|---|
| treat_MCact | 0.00197 | 0.0163$^{*}$ | 0.00502 | 0.000496 | 0.0163$^{**}$ | 0.00351 | 0.00191$^{***}$ | 0.00109 | 0.00337 | -0.000434 | -0.00109 | -0.00186 |
| (β1) | (0.00259) | (0.00944) | (0.00842) | (0.00186) | (0.00794) | (0.00757) | (0.000565) | (0.00282) | (0.00302) | (0.000666) | (0.00344) | (0.00399) |
| post_MCact | 0.0262$^{***}$ | -0.00415 | 0.0215$^{***}$ | 0.0142$^{***}$ | -0.00581$^{*}$ | 0.0161$^{***}$ | 0.00175$^{**}$ | -0.00312$^{***}$ | -0.00195 | 0.0103$^{***}$ | 0.00477$^{***}$ | 0.00728$^{***}$ |
| (β2) | (0.00396) | (0.00394) | (0.00461) | (0.00285) | (0.00331) | (0.00414) | (0.000865) | (0.00118) | (0.00165) | (0.00102) | (0.00143) | (0.00218) |
| **DiD_MCact** | **0.0217**$^{***}$ | **0.0230**$^{***}$ | **0.0228**$^{***}$ | **0.0122**$^{***}$ | **0.0128**$^{***}$ | **0.0112**$^{***}$ | **0.0103**$^{***}$ | **0.0106**$^{***}$ | **0.0117**$^{***}$ | **-0.000770** | **-0.000321** | **-0.000201** |
| (β3) | (0.00560) | (0.00356) | (0.00289) | (0.00404) | (0.00300) | (0.00260) | (0.00122) | (0.00107) | (0.00104) | (0.00144) | (0.00130) | (0.00137) |
| post_MCann | | | 0.0190$^{***}$ | | | 0.0187$^{***}$ | | | -0.000702 | | | 0.000974 |
| (β4) | | | (0.00389) | | | (0.00349) | | | (0.00139) | | | (0.00184) |
| **DiD_MCann** | | | **0.00870**$^{**}$ | | | **0.00164** | | | **0.00528**$^{***}$ | | | **0.00178** |
| (β5) | | | (0.00418) | | | (0.00376) | | | (0.00150) | | | (0.00198) |
| PlanA | | | 0.0141$^{***}$ | | | 0.0126$^{***}$ | | | -0.000431 | | | 0.00192 |
| (β6) | | | (0.00298) | | | (0.00268) | | | (0.00107) | | | (0.00141) |
| **DiD_PlanA** | | | **0.00222** | | | **-0.00179** | | | **0.00354**$^{***}$ | | | **0.000466** |
| (β7) | | | (0.00327) | | | (0.00294) | | | (0.00117) | | | (0.00155) |
| SS_AFFV | | -0.000471 | -0.000289 | | 0.00000455 | -0.0000748 | | -0.000231$^{*}$ | -0.0000111 | | -0.000245 | -0.000203 |
| (β8) | | (0.000416) | (0.000372) | | (0.000350) | (0.000335) | | (0.000125) | (0.000134) | | (0.000152) | (0.000176) |
| relLPG | | -0.0892$^{*}$ | -0.00442 | | -0.0818$^{**}$ | -0.0216 | | -0.0105 | 0.00422 | | 0.00307 | 0.0129 |
| (β9) | | (0.0461) | (0.0365) | | (0.0388) | (0.0328) | | (0.0138) | (0.0131) | | (0.0168) | (0.0173) |
| EVchgStat | | 0.000138$^{***}$ | 0.0000471$^{*}$ | | 0.0000743$^{***}$ | 0.0000137 | | 0.0000314$^{***}$ | 0.0000114 | | 0.0000326$^{***}$ | 0.0000219 |
| (β10) | | (0.0000327) | (0.0000280) | | (0.0000275) | (0.0000252) | | (0.00000979) | (0.0000101) | | (0.0000119) | (0.0000133) |
| TasaAct_ | | -0.00641$^{***}$ | -0.00509$^{***}$ | | -0.00239 | -0.00214 | | -0.00181$^{***}$ | -0.000930$^{*}$ | | -0.00221$^{***}$ | -0.00202$^{***}$ |
| (β11) | | (0.00183) | (0.00155) | | (0.00154) | (0.00139) | | (0.000547) | (0.000555) | | (0.000666) | (0.000733) |
| _cons | 0.0146$^{***}$ | 0.466$^{***}$ | 0.343$^{***}$ | 0.00731$^{***}$ | 0.177 | 0.151 | 0.00285$^{***}$ | 0.135$^{***}$ | 0.0559 | 0.00440$^{***}$ | 0.154$^{***}$ | 0.136$^{**}$ |
| | (0.00183) | (0.134) | (0.118) | (0.00132) | (0.113) | (0.106) | (0.000399) | (0.0400) | (0.0423) | (0.000471) | (0.0488) | (0.0558) |
| *N* | 122 | 122 | 122 | 122 | 122 | 122 | 122 | 122 | 122 | 122 | 122 | 122 |
| $R^2$ | 0.627 | 0.881 | 0.933 | 0.491 | 0.778 | 0.858 | 0.692 | 0.816 | 0.851 | 0.616 | 0.755 | 0.768 |
| *RMSD* | 0.01262 | 0.00728 | 0.00555 | 0.00913 | 0.00613 | 0.00499 | 0.00277 | 0.00218 | 0.00199 | 0.00326 | 0.00266 | 0.00263 |

**Table A4.** Regression results on municipal level, LEZ in combination with local subsidy scheme. Std. errors in parentheses. $^{*}$ $p < 0.1$, $^{**}$ $p < 0.05$, $^{***}$ $p < 0.01$

| **Municipality** | **(Model 1)** | **(Model 2)** | **(Model 3)** | **(Model 1)** | **(Model 2)** | **(Model 3)** | **(Model 1)** | **(Model 2)** | **(Model 3)** | **(Model 1)** | **(Model2)** | **(Model3)** |
|---|---|---|---|---|---|---|---|---|---|---|---|---|
| | | **shareAFV** | | | **shareAFFV** | | | **sharePHEV** | | | **shareEV** | |
| treat_MCact | 0.00359 | -0.000528 | -0.0112 | 0.00551$^{**}$ | -0.0105 | -0.0176 | 0.00106 | 0.00180 | 0.00406 | -0.00336$^{***}$ | 0.00725 | 0.000742 |
| (β1) | (0.00350) | (0.0151) | (0.0143) | (0.00242) | (0.0116) | (0.0110) | (0.000746) | (0.00379) | (0.00424) | (0.00118) | (0.00670) | (0.00767) |
| post_MCact | 0.0384$^{***}$ | 0.00343 | 0.0347$^{***}$ | 0.0140$^{***}$ | -0.00672 | 0.0172$^{***}$ | 0.00300$^{***}$ | -0.00355$^{**}$ | -0.00152 | 0.0210$^{***}$ | 0.0140$^{***}$ | 0.0198$^{***}$ |
| (β2) | (0.00536) | (0.00630) | (0.00782) | (0.00371) | (0.00483) | (0.00601) | (0.00113) | (0.00157) | (0.00231) | (0.00181) | (0.00279) | (0.00420) |
| **DiD_MCact** | **0.0354$^{***}$** | **0.0369$^{***}$** | **0.0391$^{***}$** | **0.0360$^{***}$** | **0.0370$^{***}$** | **0.0387$^{***}$** | **0.00885$^{***}$** | **0.00919$^{***}$** | **0.0103$^{***}$** | **-0.00910$^{***}$** | **-0.00922$^{***}$** | **-0.00987$^{***}$** |
| (β3) | (0.00758) | (0.00570) | (0.00491) | (0.00524) | (0.00437) | (0.00377) | (0.00159) | (0.00141) | (0.00141) | (0.00256) | (0.00253) | (0.00263) |
| post_MCann | | | 0.0181$^{***}$ | | | 0.0152$^{***}$ | | | -0.000170 | | | 0.00357 |
| (β4) | | | (0.00659) | | | (0.00507) | | | (0.00193) | | | (0.00354) |
| **DiD_MCann** | | | **0.0193$^{***}$** | | | **0.0151$^{***}$** | | | **0.00592$^{***}$** | | | **-0.00197** |
| (β5) | | | (0.00709) | | | (0.00546) | | | (0.00205) | | | (0.00381) |
| PlanA | | | 0.0175$^{***}$ | | | 0.0117$^{***}$ | | | -0.0000227 | | | 0.00613$^{**}$ |
| (β6) | | | (0.00506) | | | (0.00389) | | | (0.00147) | | | (0.00272) |
| **DiD_PlanA** | | | **0.0106$^{*}$** | | | **0.00830$^{*}$** | | | **0.00314$^{*}$** | | | **-0.000821** |
| (β7) | | | (0.00555) | | | (0.00427) | | | (0.00160) | | | (0.00298) |
| SS_AFFV | | -0.00160$^{**}$ | -0.00108$^{*}$ | | -0.00161$^{***}$ | -0.00114$^{**}$ | | -0.000210 | 0.0000315 | | 0.000181 | -0.0000178 |
| (β8) | | (0.000666) | (0.000632) | | (0.000510) | (0.000486) | | (0.000166) | (0.000185) | | (0.000295) | (0.000339) |
| relLPG | | -0.108 | 0.0204 | | -0.100$^{*}$ | -0.00214 | | -0.0156 | 0.00171 | | 0.00719 | 0.0222 |
| (β9) | | (0.0738) | (0.0619) | | (0.0565) | (0.0476) | | (0.0184) | (0.0181) | | (0.0327) | (0.0333) |
| EVchgStat | | 0.000225$^{***}$ | 0.0000906$^{*}$ | | 0.000179$^{***}$ | 0.0000719$^{*}$ | | 0.0000372$^{***}$ | 0.0000130 | | 0.00000939 | 0.00000562 |
| (β10) | | (0.0000524) | (0.0000476) | | (0.0000401) | (0.0000366) | | (0.0000129) | (0.0000137) | | (0.0000232) | (0.0000255) |
| TasaAct_ | | -0.0126$^{***}$ | -0.00991$^{***}$ | | -0.00952$^{***}$ | -0.00714$^{***}$ | | -0.00160$^{**}$ | -0.000659 | | -0.00171 | -0.00239$^{*}$ |
| (β11) | | (0.00292) | (0.00262) | | (0.00224) | (0.00202) | | (0.000741) | (0.000778) | | (0.00130) | (0.00141) |
| _cons | 0.0180$^{***}$ | 0.955$^{***}$ | 0.699$^{***}$ | 0.00545$^{***}$ | 0.761$^{***}$ | 0.541$^{***}$ | 0.00449$^{***}$ | 0.122$^{**}$ | 0.0369 | 0.00847$^{***}$ | 0.0885 | 0.143 |
| (β0) | (0.00247) | (0.214) | (0.200) | (0.00171) | (0.164) | (0.154) | (0.000539) | (0.0544) | (0.0594) | (0.000836) | (0.0950) | (0.107) |
| *N* | 122 | 122 | 122 | 122 | 122 | 122 | 118 | 118 | 118 | 122 | 122 | 122 |
| $R^2$ | 0.682 | 0.858 | 0.911 | 0.665 | 0.817 | 0.883 | 0.549 | 0.719 | 0.759 | 0.632 | 0.717 | 0.738 |
| *RMSD* | 0.01714 | 0.01165 | 0.00942 | 0.01186 | 0.00893 | 0.00724 | 0.00358 | 0.00287 | 0.00271 | 0.00579 | 0.00517 | 0.00506 |

**Table A5.** Regression results on province level, LEZ without local subsidy scheme. Standard errors in parentheses. $^{*}$ $p < 0.1$, $^{**}$ $p < 0.05$, $^{***}$ $p < 0.01$

| Province | (Model 1) | (Model 2) | (Model 3) | (Model 1) | (Model 2) | (Model 3) | (Model 1) | (Model 2) | (Model 3) | (Model 1) | (Model2) | (Model3) |
|---|---|---|---|---|---|---|---|---|---|---|---|---|
| | shareAFV_noMUS | | | shareAFFV_noMUS | | | sharePHEV_noMUS | | | shareEV_noMUS | | |
| treat_MCact | 0.00197 | 0.0163$^{*}$ | 0.00487 | 0.000496 | 0.0162$^{**}$ | 0.00342 | 0.00191$^{***}$ | 0.00113 | 0.00325 | -0.000434 | -0.00103 | -0.00180 |
| (β1) | (0.00258) | (0.00939) | (0.00828) | (0.00184) | (0.00773) | (0.00724) | (0.000565) | (0.00283) | (0.00303) | (0.000665) | (0.00343) | (0.00397) |
| post_MCact | 0.0262$^{***}$ | -0.00418 | 0.0217$^{***}$ | 0.0142$^{***}$ | -0.00589$^{*}$ | 0.0161$^{***}$ | 0.00175$^{**}$ | -0.00304$^{**}$ | -0.00170 | 0.0103$^{***}$ | 0.00476$^{***}$ | 0.00730$^{***}$ |
| (β2) | (0.00395) | (0.00392) | (0.00453) | (0.00282) | (0.00322) | (0.00396) | (0.000865) | (0.00118) | (0.00166) | (0.00102) | (0.00143) | (0.00217) |
| **DiD_MCact** | 0.0169$^{***}$ | 0.0182$^{***}$ | 0.0179$^{***}$ | 0.00954$^{**}$ | 0.0102$^{***}$ | 0.00860$^{***}$ | 0.00907$^{***}$ | 0.00934$^{***}$ | 0.0105$^{***}$ | -0.00176 | -0.00129 | -0.00117 |
| (β3) | (0.00558) | (0.00354) | (0.00284) | (0.00398) | (0.00292) | (0.00248) | (0.00122) | (0.00107) | (0.00104) | (0.00144) | (0.00129) | (0.00136) |
| post_MCann | | | 0.0192$^{***}$ | | | 0.0187$^{***}$ | | | -0.000534 | | | 0.000990 |
| (β4) | | | (0.00382) | | | (0.00334) | | | (0.00140) | | | (0.00183) |
| **DiD_MCann** | | | 0.00871$^{**}$ | | | 0.00168 | | | 0.00522$^{***}$ | | | 0.00181 |
| (β5) | | | (0.00411) | | | (0.00359) | | | (0.00150) | | | (0.00197) |
| PlanA | | | 0.0143$^{***}$ | | | 0.0126$^{***}$ | | | -0.000313 | | | 0.00193 |
| (β6) | | | (0.00293) | | | (0.00256) | | | (0.00107) | | | (0.00141) |
| **DiD_PlanA** | | | 0.00224 | | | -0.00176 | | | 0.00351$^{***}$ | | | 0.000486 |
| (β7) | | | (0.00322) | | | (0.00281) | | | (0.00118) | | | (0.00154) |
| SS_AFFV | | -0.000466 | -0.000285 | | 0.00000366 | -0.0000740 | | -0.000229$^{*}$ | -0.0000127 | | -0.000241 | -0.000198 |
| (β8) | | (0.000414) | (0.000366) | | (0.000341) | (0.000320) | | (0.000125) | (0.000134) | | (0.000151) | (0.000176) |
| relLPG | | -0.0865$^{*}$ | -0.000989 | | -0.0818$^{**}$ | -0.0214 | | -0.00900 | 0.00614 | | 0.00429 | 0.0143 |
| (β9) | | (0.0459) | (0.0359) | | (0.0377) | (0.0314) | | (0.0138) | (0.0131) | | (0.0168) | (0.0172) |
| EVchgStat | | 0.000138$^{***}$ | 0.0000459$^{*}$ | | 0.0000747$^{***}$ | 0.0000139 | | 0.0000308$^{***}$ | 0.0000106 | | 0.0000323$^{***}$ | 0.0000214 |
| (β10) | | (0.0000326) | (0.0000276) | | (0.0000268) | (0.0000241) | | (0.00000983) | (0.0000101) | | (0.0000119) | (0.0000132) |
| TasaAct_ | | -0.00636$^{***}$ | -0.00504$^{***}$ | | -0.00231 | -0.00205 | | -0.00185$^{***}$ | -0.000981$^{*}$ | | -0.00220$^{***}$ | -0.00201$^{***}$ |
| (β11) | | (0.00182) | (0.00152) | | (0.00150) | (0.00133) | | (0.000549) | (0.000556) | | (0.000664) | (0.000730) |
| _cons | 0.0146$^{***}$ | 0.461$^{***}$ | 0.338$^{***}$ | 0.00731$^{***}$ | 0.172 | 0.145 | 0.00285$^{***}$ | 0.136$^{***}$ | 0.0583 | 0.00440$^{***}$ | 0.153$^{***}$ | 0.134$^{**}$ |
| (β0) | (0.00182) | (0.133) | (0.116) | (0.00130) | (0.110) | (0.101) | (0.000399) | (0.0402) | (0.0424) | (0.000470) | (0.0486) | (0.0556) |
| *N* | 122 | 122 | 122 | 122 | 122 | 122 | 122 | 122 | 122 | 122 | 122 | 122 |
| $R^2$ | 0.589 | 0.869 | 0.928 | 0.455 | 0.770 | 0.858 | 0.650 | 0.789 | 0.830 | 0.596 | 0.742 | 0.756 |
| *RMSD* | 0.01262 | 0.00724 | 0.00546 | 0.00901 | 0.00596 | 0.00477 | 0.00277 | 0.00219 | 0.0020 | 0.00326 | 0.00265 | 0.00262 |

**Table A6.** Regression results on municipal level, LEZ without local subsidy scheme. Standard errors in parentheses. $^{*}$ $p$ < 0.1, $^{**}$ $p$ < 0.05, $^{***}$ $p$ < 0.01

| Municipality | (Model 1) | (Model 2) | (Model 3) | (Model 1) | (Model 2) | (Model 3) | (Model 1) | (Model 2) | (Model 3) | (Model 1) | (Model2) | (Model3) |
|---|---|---|---|---|---|---|---|---|---|---|---|---|
| | shareAFV_noMUS | | | shareAFFV_noMUS | | | sharePHEV_noMUS | | | shareEV_noMUS | | |
| treat_MCact | 0.00797$^{**}$ | -0.00178 | -0.0124 | 0.00817$^{***}$ | -0.0118 | -0.0186 | 0.00106 | 0.00187 | 0.00401 | -0.00336$^{***}$ | 0.00735 | 0.000872 |
| (β1) | (0.00366) | (0.0160) | (0.0155) | (0.00252) | (0.0119) | (0.0114) | (0.000734) | (0.00370) | (0.00411) | (0.00119) | (0.00673) | (0.00771) |
| post_MCact | 0.0384$^{***}$ | 0.00382 | 0.0351$^{***}$ | 0.0140$^{***}$ | -0.00648 | 0.0173$^{***}$ | 0.00300$^{***}$ | -0.00349$^{**}$ | -0.00134 | 0.0210$^{***}$ | 0.0139$^{***}$ | 0.0197$^{***}$ |
| (β2) | (0.00527) | (0.00648) | (0.00803) | (0.00363) | (0.00484) | (0.00592) | (0.00111) | (0.00153) | (0.00224) | (0.00182) | (0.00281) | (0.00422) |
| **DiD_MCact** | 0.0235$^{***}$ | 0.0286$^{***}$ | 0.0311$^{***}$ | 0.0285$^{***}$ | 0.0316$^{***}$ | 0.0336$^{***}$ | 0.00747$^{***}$ | 0.00780$^{***}$ | 0.00890$^{***}$ | -0.0104$^{***}$ | -0.0105$^{***}$ | -0.0111$^{***}$ |
| (β3) | (0.00756) | (0.00586) | (0.00509) | (0.00521) | (0.00438) | (0.00376) | (0.00156) | (0.00137) | (0.00137) | (0.00257) | (0.00254) | (0.00265) |
| post_MCann | | | 0.0188$^{***}$ | | | 0.0155$^{***}$ | | | -0.0000558 | | | 0.00356 |
| (β4) | | | (0.00694) | | | (0.00512) | | | (0.00187) | | | (0.00356) |
| **DiD_MCann** | | | 0.0182$^{**}$ | | | 0.0145$^{**}$ | | | 0.00590$^{***}$ | | | -0.00191 |
| (β5) | | | (0.00755) | | | (0.00557) | | | (0.00198) | | | (0.00383) |
| PlanA | | | 0.0179$^{***}$ | | | 0.0118$^{***}$ | | | 0.0000562 | | | 0.00612$^{**}$ |
| (β6) | | | (0.00526) | | | (0.00387) | | | (0.00142) | | | (0.00273) |
| **DiD_PlanA** | | | 0.00989$^{*}$ | | | 0.00797$^{*}$ | | | 0.00313$^{**}$ | | | -0.000782 |
| (β7) | | | (0.00584) | | | (0.00431) | | | (0.00155) | | | (0.00299) |
| SS_AFFV | | -0.00167$^{**}$ | -0.00115$^{*}$ | | -0.00168$^{***}$ | -0.00121$^{**}$ | | -0.000207 | 0.0000325 | | 0.000187 | -0.00000923 |
| (β8) | | (0.000703) | (0.000685) | | (0.000525) | (0.000505) | | (0.000162) | (0.000179) | | (0.000297) | (0.000341) |
| relLPG | | -0.102 | 0.0338 | | -0.101 | 0.00276 | | -0.0137 | 0.00393 | | 0.00892 | 0.0241 |
| (β9) | | (0.0817) | (0.0686) | | (0.0610) | (0.0506) | | (0.0179) | (0.0175) | | (0.0329) | (0.0334) |
| EVchgStat | | 0.000228$^{***}$ | 0.0000924$^{*}$ | | 0.000183$^{***}$ | 0.0000752$^{**}$ | | 0.0000365$^{***}$ | 0.0000121 | | 0.00000894 | 0.00000494 |
| (β10) | | (0.0000556) | (0.0000511) | | (0.0000415) | (0.0000377) | | (0.0000126) | (0.0000133) | | (0.0000233) | (0.0000257) |
| TasaAct_ | | -0.0124$^{***}$ | -0.0102$^{***}$ | | -0.00954$^{***}$ | -0.00744$^{***}$ | | -0.00165$^{**}$ | -0.000707 | | -0.00169 | -0.00236$^{*}$ |
| (β11) | | (0.00325) | (0.00284) | | (0.00243) | (0.00210) | | (0.000722) | (0.000754) | | (0.00130) | (0.00142) |
| _cons | 0.0180$^{***}$ | 0.947$^{***}$ | 0.716$^{***}$ | 0.00545$^{***}$ | 0.769$^{***}$ | 0.563$^{***}$ | 0.00449$^{***}$ | 0.124$^{**}$ | 0.0388 | 0.00847$^{***}$ | 0.0860 | 0.139 |
| (β0) | (0.00243) | (0.232) | (0.213) | (0.00168) | (0.174) | (0.157) | (0.000530) | (0.0531) | (0.0576) | (0.000839) | (0.0954) | (0.108) |
| *N* | 112 | 112 | 112 | 112 | 112 | 112 | 118 | 118 | 118 | 122 | 122 | 122 |
| $R^2$ | 0.657 | 0.836 | 0.897 | 0.647 | 0.801 | 0.878 | 0.500 | 0.694 | 0.741 | 0.624 | 0.710 | 0.732 |
| *RMSD* | 0.01687 | 0.0119 | 0.00963 | 0.01162 | 0.00889 | 0.0071 | 0.00351 | 0.0028 | 0.00263 | 0.00581 | 0.00519 | 0.00508 |

## Acknowledgements

This work was funded by the European Union's Horizon 2020 Research and Innovation Programme under the Marie Skłodowska-Curie Grant Agreement No. 754382. The authors would like to thank Electromaps for their courtesy of sharing data on EV charging stations.

## Author Contributions

JFP conceived and performed the research and wrote the manuscript, JMAM revised the econometric models and -results and reviewed the manuscript, MBC provided input during the conceptualization and design phase and reviewed the manuscript.

## Disclaimer

The content of this article does not reflect the official opinion of the European Union. Responsibility for the information and views expressed herein lies entirely with the authors

## Supplementary Material

Supplementary Material to this article and the underlying data for import and reproduction plus corresponding summary statistics as dataset in Excel-format are provided on Zenodo (DOI: 10.5281/zenodo.3948364) and the Metabolism of Cities Data Hub (https://data.metabolismofcities.org/dashboards/madrid/datasets/ )